\documentstyle[prl,aps,psfig,floats,preprint]{revtex}

\begin{document}
\def\beq{\begin{equation}}
\def\eeq{\end{equation}}
\def\ber{\begin{eqnarray}}
\def\eer{\end{eqnarray}}
\def\apj{{Astroph.\@ J.\ }}
\def\mn{{Mon.\@ Not.@ Roy.\@ Ast.\@ Soc.\ }}
\def\asta{{Astron.\@ Astrophys.\ }}
\def\aj{{Astron.\@ J.\ }}
\def\prl{{Phys.\@ Rev.\@ Lett.\ }}
\def\prd{{Phys.\@ Rev.\@ D\ }}
\def\nucp{{Nucl.\@ Phys.\ }}
\def\nat{{Nature\ }}
\def\plb {{Phys.\@ Lett.\@ B\ }}
\def \jetpl {JETP Lett.\ }
\def\etal{{\it et al.}}
\def\ie {{\it ie}}

\title{New Cosmological Singularities in Braneworld Models}

\author{Yuri Shtanov$^a$ and Varun Sahni$^b$}
\address{$^a$Bogolyubov Institute for Theoretical Physics, Kiev 03143,
Ukraine \\
$^b$Inter-University Centre for Astronomy and Astrophysics, Post Bag 4,
Ganeshkhind, Pune 411~007, India}

\maketitle

\thispagestyle{empty}

\sloppy

\begin{abstract}
Higher-dimensional braneworld  models which contain both bulk and brane
curvature terms in the action admit cosmological singularities of rather
unusual form and nature. These `quiescent' singularities, which can occur both
during the contracting as well as the expanding phase, are characterised by the
fact that while the matter density and Hubble parameter remain finite, all
higher derivatives of the scale factor ($\stackrel{..}{a}$, $\stackrel{...}{a}$
etc.) diverge as the cosmological singularity is approached. The singularities
are the result of the embedding of the (3+1)-dimensional brane in the bulk and
can exist even in an empty homogeneous and isotropic (FRW) universe. The
possibility that the present universe may expand into a singular state is
discussed.

\end{abstract}

%\pacs{PACS number(s): 04.50.+h, 98.80.Hw}

\bigskip \medskip

In this letter, we would like to draw attention to the fact that braneworld
theory admits cosmological singularities of very unusual form and nature.  The
theory we work with in this paper is described by the action
\begin{equation} \label{action}
S = M^3 \sum_i \left[\int_{\rm bulk} \left( {\cal R} - 2 \Lambda_i \right) - 2
\int_{\rm brane} K \right] + \int_{\rm brane} \left( m^2 R - 2 \sigma \right) +
\int_{\rm brane} L \left( h_{\alpha\beta}, \phi \right) \, ,
\end{equation}
the notation of which is standard. We use the signature and sign conventions of
\cite{Wald}. The sum in (\ref{action}) is taken over the bulk components
bounded by branes, and $\Lambda_i$ is the cosmological constant in the $i$th
bulk component. Note that we have included the curvature term in the action for
the brane with the coefficient $m^2$. Such a term generically arises when one
considers quantum effects generated by matter fields residing on the brane
\cite{CHS,DGP}, in the spirit of the idea of the induced effective action
\cite{Sakharov,BD}; its inclusion in the action is therefore mandatory. The
lagrangian $L \left( h_{\alpha\beta}, \phi \right)$ corresponds to the presence
of matter fields $\phi$ on the brane interacting with the induced metric
$h_{\alpha\beta}$ and describes their dynamics. The extrinsic curvature
$K_{\alpha\beta}$ on both sides of the brane is defined with respect to the
inner normal $n^a$, as in \cite{Shtanov1,Shtanov2,ss02}.

Let us first consider the case with $Z_2$ symmetry of reflection, which
requires equal cosmological constants on the two sides of the brane: $\Lambda_1
= \Lambda_2 = \Lambda$. The corresponding cosmological equation of theory
(\ref{action}) was derived in \cite{CHS} and has the form
\begin{equation} \label{solution}
H^2 + {\kappa \over a^2} = {\rho + \sigma \over 3 m^2} + {2 \over \ell^2}
\left[1 \pm \sqrt{1 + \ell^2 \left({\rho + \sigma \over 3 m^2} - {\Lambda \over
6} - {C \over a^4} \right)} \right] \, ,
\end{equation}
where the integration constant $C$ corresponds to the presence of a black hole
in the five-dimensional bulk solution, and the term $C/a^4$ (occasionally
referred to as `dark radiation') arises due to the projection of the bulk
gravitational degrees of freedom onto the brane.  The length scale $\ell$ is
defined as
\begin{equation}
\ell = \frac{m^2}{M^3} \, .
\end{equation}

The new singularities that we are going to discuss in this paper are connected
with the fact that the expression under the square root of (\ref{solution})
{\em turns to zero\/} at some point during evolution, so that solutions of the
cosmological equations {\em cannot be continued\/} beyond this point.  There
are essentially two types of `quiescent' singularities displaying this
behaviour:

A {\bf type 1} singularity (S1) is essentially induced by the presence of the
`dark radiation' term under the square root of (\ref{solution}) and arises in
either of the following two cases:

\begin{itemize}

\item
$C > 0$ and the density of matter increases {\em slower\/} than $a^{-4}$ as $a
\to 0$. Such singularities occur if the universe is filled with matter having
equation of state $P/\rho < 1/3$, an example is provided by pressureless matter
(dust) for which $\rho \propto a^{-3}$.  A special case is an empty universe
($\rho = 0$).

\item
The energy density of the universe is radiation-dominated so that $\rho =
\rho_0  / a^4$ and $C > \rho_0 $.

\end{itemize}

The singularities discussed above can take place either in the past of an
expanding universe or in the future of a collapsing one.

A {\bf type 2} singularity (S2) arises if
\begin{equation} \label{ii}
\ell^2 \left({\sigma \over 3 m^2} - {\Lambda \over 6} \right) < - 1 \, .
\end{equation}

In this case, it is important to note that the combination $\rho / 3 m^2 - C
/a^4$ decreases monotonically as the universe expands. The expression under the
square root of (\ref{solution}) can therefore become zero at suitably late
times, in which case the cosmological solution cannot be extended beyond this
time. S2 is even more interesting than S1 since: (i)  it can occur during the
late time expansion of the universe; (ii) it can occur {\em even if\/} dark
radiation is entirely absent ($C = 0$).

\begin{figure}[tbh!]
\centerline{ \psfig{figure=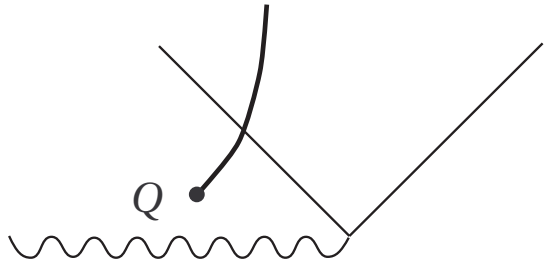,width=0.5\textwidth,angle=0} }
\bigskip
\caption{\small Conformal diagram showing the trajectory of a spatially
spherical braneworld embedded in the five-dimensional Schwarzschild space-time.
The trajectory is not smoothly extendable beyond the point $Q$.}
\label{fig:brane}
\end{figure}

For both S1 \& S2, the scale factor $a(t)$ and its first time derivative remain
finite, while all the higher time derivatives of $a(t)$ tend to infinity as the
singularity is approached. As an example consider a type 2 singularity with $C
= 0$, for which
\begin{equation}
\frac{d^na}{dt^n} = {\cal O} \left( \left[ \rho(t) - \alpha \right]^{3/2-n}
\right), \quad n \geq 2 \, ,
\end{equation}
as $\rho(t) \to \alpha = \Lambda m^2/2 - \sigma - 3m^2/\ell^2$. We therefore find
that the scalar curvature $R \to \infty$ near the singularity, while the energy
density and pressure {\em remain finite}. Although this situation is quite
unusual from the viewpoint of the intrinsic dynamics on the brane, it becomes
comprehensible when one considers the embedding of the brane in the bulk. It is
well known that the cosmological braneworld under consideration can be
isometrically embedded in the five-dimensional solution of the vacuum Einstein
equations described by the metric
\begin{equation} \label{ads}
ds^2 = - f_\kappa (r) dt^2 + {dr^2 \over f_\kappa (r)} + r^2 d\Omega_\kappa \,
,
\end{equation}
where
\begin{equation}
f_\kappa (r) = \kappa - {\Lambda \over 6} r^2 - {C \over r^2} \, ,
\end{equation}
and $d\Omega_\kappa$ is the metric of the three-dimensional Euclidean space,
sphere or pseudosphere corresponding to the value of $\kappa = 0, \pm 1$.  The
embedding of the brane is defined by the function $r = a(t)$, and one can then
proceed to define evolution in terms of the proper cosmological time $\tau$ of
the induced metric on the brane. The cosmological singularity under
consideration is connected with the fact that the brane embedding is not
extendable beyond some moment of time $t$ because the function $a(t)$ that
defines the embedding cannot be smoothly continued beyond this point (see
Fig.~\ref{fig:brane}).

\begin{figure}[tbh!]
\centerline{ \psfig{figure=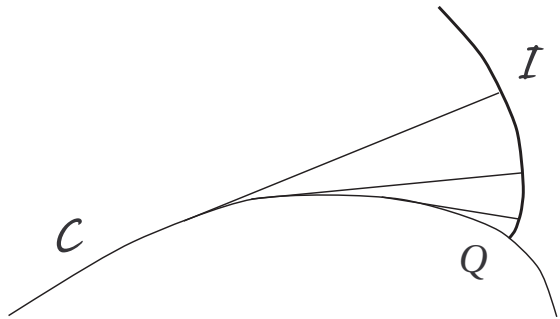,width=0.5\textwidth,angle=0} }
\bigskip
\caption{\small The involute ${\cal I}$ of a planar curve ${\cal C}$ is not
smoothly extendable beyond the starting point $Q$.} \label{fig:involute}
\end{figure}

This specific feature of the brane embedding can be compared to the behaviour
of the involute of a planar curve.  The involute ${\cal I}$ of a convex planar
curve ${\cal C}$ is a line which intersects the tangent lines of ${\cal C}$
orthogonally \cite{lip}. ${\cal I}$ can be visualised as the trajectory
described by the end of a strained thread winding up from ${\cal C}$ (see
Fig.~\ref{fig:involute}).  The involute of a typical curve is sharp at the
starting point $Q$ so that it is not smoothly extendable beyond the point $Q$.

This analogy can be traced further.  Note that the evolution of the brane in
theory (\ref{action}) is described by the following well-known equation:
\begin{equation} \label{brane}
m^2 G_{\alpha\beta} + \sigma h_{\alpha\beta} = T_{\alpha\beta} + M^3 \sum
\left( K_{\alpha\beta} - h_{\alpha\beta} K \right) \, ,
\end{equation}
where the extrinsic curvature is summed over both sides of the brane.  One can
see that it is the influence of the sum of the extrinsic curvatures on the
right-hand side that leads to the singularities under investigation so that the
singularity of the Einstein tensor $G_{\alpha\beta}$ is accompanied by the
singularity of the extrinsic curvature $K_{\alpha\beta}$, while the induced
metric $h_{\alpha\beta}$ and the stress-energy tensor $T_{\alpha\beta}$ on the
brane remain finite. Quite similarly, the involute of a curve is defined
through the extrinsic curvature of its embedding in the plane, as is clear from
Fig.~\ref{fig:involute}, and its singularity at the point $Q$ is connected with
the fact that the extrinsic curvature diverges at this point. Specifically, the
parametric equation for the involute ${\bf x}_* (s)$, $s \ge 0$, in Cartesian
coordinates on the plane can be written as \cite{lip}
\begin{equation}
{\bf x}_* (s) = {\bf x} (q - s) + s \cdot {\bf x}' (q - s) \, ,
\end{equation}
where ${\bf x} (s)$ is the curve ${\cal C}$ parametrized by the natural
parameter $s$, and ${\bf x} (q) = {\bf x}_* (0)$ is the coordinate of the
starting point $Q$ of the involute.  The extrinsic curvature of the involute is
\begin{equation}
k (s) = \frac{1}{s} \, ,
\end{equation}
which diverges at the starting point $Q$ corresponding to $s = 0$.

One should also highlight an important difference between the 1D and 4D
embeddings: the involute being one-dimensional, a singularity in its extrinsic
curvature does not lead to a singularity in its intrinsic geometry. As we have
seen, this is not the case with the brane for which the extrinsic and intrinsic
curvatures are related through (\ref{brane}), so that a singularity in
$K_{\alpha\beta}$ is reflected in a singularity in $G_{\alpha\beta}$.

Interestingly, an S2 singularity can arise in the distant future of a universe
resembling our own~! To illustrate this we rewrite Eq.~(\ref{solution}) (with
$C = 0$) as \cite{ss02}
\begin{equation} \label{hubble1}
{H^2(z) \over H_0^2} = \Omega_{\rm m} (1\!+\!z)^3 + \Omega_\kappa (1\!+\!z)^2 +
\Omega_\sigma + \underline{2 \Omega_\ell \pm 2 \sqrt{\Omega_\ell}\,
\sqrt{\Omega_{\rm m} (1\!+\!z )^3 + \Omega_\sigma + \Omega_\ell +
\Omega_{\Lambda}}} \, ,
\end{equation}
The underlining emphasises those terms which cause the braneworld to differ
from its general-relativistic counterpart. For simplicity, we shall only
discuss the solution corresponding to the `+' sign in (\ref{hubble1}) (called
BRANE2 in \cite{ss02}). In this case, it can be shown that our model satisfies
the constraint equation
\begin{equation} \label{omega-r2}
\Omega_{\rm m} + \Omega_\kappa + \Omega_\sigma + \underline{2
\sqrt{\Omega_\ell}\, \sqrt{1 - \Omega_\kappa + \Omega_\Lambda}} = 1 \, ,
\end{equation}
where
\begin{equation} \label{omegas}
\Omega_{\rm m} =  {\rho_0 \over 3 m^2 H_0^2} \, , \quad \Omega_\kappa = -
{\kappa \over a_0^2 H_0^2} \, , \quad \Omega_\sigma = {\sigma \over 3 m^2
H_0^2} \, , \quad \Omega_\ell = {1 \over \ell^2 H_0^2} \, , \quad
\Omega_\Lambda = - {\Lambda \over 6 H_0^2} \, ,
\end{equation}
and the subscript `{\small 0}' refers to the present value of
the various cosmological
quantities.

\begin{figure}[tbh!]
\centerline{ \psfig{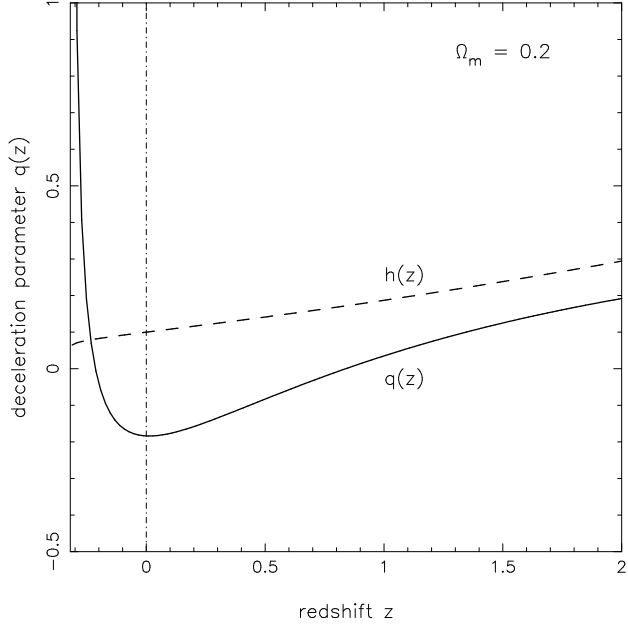} }
\bigskip
\caption{\small The deceleration parameter (solid line) is shown for a
braneworld model with $\Omega_{\rm m} = 0.2$, $\Omega_\ell = 0.4$,
$\Omega_\Lambda = \Omega_\kappa = 0$, and $\Omega_\sigma$ determined from
(\ref{omega-r2}). We find that $q(z) \to 0.5$ for $z \gg 1$ while $q(z) \to
\infty$ as $z \to -0.312779$... Currently $q_0 < 0$, which indicates that the
universe is accelerating. Also shown is the dimensionless Hubble parameter
$h(z) = 0.1\times H(z)/H_0$ (dashed line) for this model. The vertical line at
$z = 0$ shows the present epoch.} \label{fig:decel}
\end{figure}

Inequality (\ref{ii}) now becomes
\begin{equation} \label{constraint}
\Omega_\sigma + \Omega_\ell + \Omega_\Lambda  < 0 \, ,
\end{equation}
and the limiting redshift, $z_s = a_0/a(z_s) - 1$, at which the braneworld
becomes singular is given by
\begin{equation}
z_s = \left (-\frac{\Omega_\sigma + \Omega_\ell + \Omega_\Lambda} {\Omega_{\rm
m}}\right )^{1/3} - 1 \, .
\end{equation}
The time of occurance of the singularity (measured from the present moment)
can easily be determined from
\beq
T_s = t(z=z_s) - t(z=0) = \int_{z_s}^0\frac{dz}{(1+z)H(z)},
\eeq
where $H(z)$ is given by (\ref{hubble1}) (see also \cite{star99}).
In Fig.~\ref{fig:decel} we show a specific braneworld model having $\Omega_{\rm
m} = 0.2$, $\Omega_\ell = 0.4$, $\Omega_\Lambda = \Omega_\kappa = 0$. In
keeping with observations of high redshift supernovae our model universe is
currently accelerating \cite{sn}, but will become singular at $z_s \simeq -0.3
\Rightarrow a(z_s) \simeq 1.4\times a_0$, 
i.e. after $T_s \simeq 4.5~ h_{100}^{-1}$ Gyr ($h_{100} = H_0/100$ km/sec/Mpc).
Figure~\ref{fig:decel} demonstrates that
the deceleration parameter becomes singular as $z_s$ is approached:
\begin{equation}
q = -\frac{\ddot a}{aH^2} \equiv \frac{H'}{H} (1+z) - 1 \, ; \quad \lim_{z \to
z_s} q(z) \to \infty \, ,
\end{equation}
while the Hubble parameter remains finite:
\begin{equation}
{H^2(z_s) \over H_0^2} = \Omega_\ell - \Omega_\Lambda \, .
\end{equation}
It should be noted that, for a subset of parameter values, inequality
(\ref{constraint}) can be satisfied simultaneously with $\Omega_\Lambda >
\Omega_\ell$\,. In these models, the universe will recollapse (under the
influence of the negative brane tension) before the S2 singularity is reached.
(The marginal case $\Omega_\Lambda = \Omega_\ell$ corresponds to the Hubble
parameter vanishing {\em at\/} the singularity.)

Now let us briefly consider the general braneworld without $Z_2$ symmetry. In
this case, theory (\ref{action}) leads to the following general cosmological
equation for the brane embedded into the five-dimensional bulk \cite{Shtanov1}:
\begin{equation} \begin{array}{lcl} \label{cosmo}
\displaystyle m^4 \left( H^2 + {\kappa \over a^2} - {\rho + \sigma \over 3 m^2}
\right)^2 &=& \displaystyle 4 M^6  \left(H^2 + {\kappa \over a^2} - {\Lambda_1
+ \Lambda_2 \over 12} - {C \over a^4} \right)  \medskip \\ &-& \displaystyle
{M^{12} \over 36 m^4} \left[ {\Lambda_1 - \Lambda_2 + E/a^4 \over H^2 + \kappa
/ a^2 - (\rho + \sigma) / 3m^2 } \right]^2 ,
\end{array}
\end{equation}
where $C$ has the same meaning as in (\ref{solution}), $E$ is another arbitrary
integration constant, and $\Lambda_1$ and $\Lambda_2$ are the cosmological
constants on the two sides of the brane. In this paper, we restrict ourselves
to the situation where $\Lambda_1 = \Lambda_2 = \Lambda$, but $E \ne 0$. One
should note that (\ref{cosmo}) reduces to (\ref{solution}) if $\Lambda_1 =
\Lambda_2 = \Lambda$ and $E = 0$ \cite{CHS,Shtanov1}.

It is easy to see that singularity S2 is always present in the past of the
expanding brane. The reason for this rests in the {\em negative\/} character of
the last term on the right-hand side of Eq.~(\ref{cosmo}), which rapidly grows
by absolute value as $a \to 0$, while the left-hand side of this equation is
constrained to remain positive. In the case of an expanding brane (assuming
$\Lambda_1 = \Lambda_2 = \Lambda$), the last term on the right-hand side of
(\ref{cosmo}) rapidly decays and becomes unimportant. Therefore, provided
(\ref{ii}) is satisfied, the expanding universe will encounter an S2
singularity in its future.

We note that all the singularities described above are absent in the limit $m
\to 0$ which is frequently discussed in the literature.  Indeed, in this limit,
Eq.~(\ref{cosmo}) takes the form
\begin{equation}
H^2 + {\kappa \over a^2} = {\Lambda_1 + \Lambda_2 \over 12} + {C \over a^4} +
{(\rho + \sigma)^2 \over 36 M^6} + {M^6 \over 16} \left({\Lambda_1 - \Lambda_2
+ E / a^4 \over \rho + \sigma} \right)^2 \, ,
\end{equation}
which only admits cosmological singularities associated with an infinite
density of matter and dark radiation ($C/a^4$).  These singularities are
reached as $a \to 0$. (In general relativity, homogeneous and isotropic
space-times generically admit only infinite-density singularities, whereas
anisotropic space-times can be empty yet singular \cite{mac}.)

Singularites of the kind discussed in this paper occur when the original
equations of motion are non-linear with respect to the highest derivative. They
have earlier been discussed in the context of Einstein gravity with the
conformal anomaly \cite{wfhh}. (This result is not surprising in view of the
formal similarity between braneworld theory and GR-based models with the
conformal anomaly, discussed in \cite{ss02}.) `Determinant singularities'
having a similar structure and properties are known to arise in the anisotropic
Bianchi~I model containing a dilaton coupled to a Gauss--Bonnet term in the
action \cite{topor} (the general theory of such `determinant singularities' is
discussed in \cite{sing}). We should emphasise that both the past (S1) and
future (S1, S2) singularities in the braneworld scenario occur for a wide range
in parameter space and might provide an interesting alternative to the `big
bang'/`big crunch' singularities of general relativity. In addition, the
universe of Fig.~\ref{fig:decel} which terminates at the S2 singularity with
$\rho_m \to$ constant as $t \to t_s$, provides an interesting contrast to the
bleak finale presented by quintessence or cosmological-constant dominated
models, in which $\rho_m \to 0$ as $t \to \infty$.

Finally, we would like to draw the reader's attention to issues which we feel
require further investigation. First of all, the singularities discussed in
this paper occur for the very special class of homogeneous and isotropic
spaces. The singularities in this case are characterized by the property that
the energy density remains bounded while the curvature blows up as one
approaches the space-time singularity. It is open to question whether
singularities of this kind will persist in space-times having less symmetry,
including the more general anisotropic and inhomogeneous case. Secondly, as the
singularity is approached and the scalar curvature of the induced metric on the
brane diverges, higher-order derivative terms in the effective action for the
brane become important. Taking these into account may qualitatively modify the
behaviour of solutions as is the case in general relativity.  Thirdly,
back-reaction associated with quantum effects such as particle production and
vacuum polarization may influence the behaviour of cosmological solutions in
the vicinty of the singularity. We hope to return to these issues in our
subsequent work.

\bigskip

{\em Acknowledgments\/}: The authors acknowledge useful discussions with Alexei
Starobinsky and support from the Indo-Ukrainian program of cooperation in
science and technology. The work of Yu.~S.\@ was also supported in part by the
INTAS grant for project No.~2000-334.

\end{document}